\documentclass[%
 aps,%
 prb,%
 amsmath,%
 amssymb,%
 floatfix,%
 a4paper,%
 showpacs,%
 twocolumn%
]{revtex4-1}

\usepackage{graphicx}
\usepackage{dcolumn}
\usepackage{bm}
\usepackage[caption=false]{subfig}
\usepackage{xcolor}

\begin{document}




\title{Segregation, precipitation, and
  $\boldsymbol\alpha$-$\boldsymbol\alpha$' phase separation in Fe-Cr
  alloys: a multi-scale modelling approach}

\author{A. Kuronen}

\affiliation{Department of Physics, University of Helsinki, P.O.Box
  43, FI-00014 Helsinki, Finland}

\author{M. H. Heinonen}
\author{S. Granroth}
\author{R. E. Per\"al\"a}
\author{T. Kilpi}
\author{P. Laukkanen}
\author{J. L{\aa}ng}
\author{J. Dahl}
\author{M. P. J. Punkkinen}
\author{K. Kokko} \email{kalevi.kokko@utu.fi}

\affiliation{Department of Physics and Astronomy, University of Turku,
  FI-20014 Turku, Finland} \affiliation{Turku University Centre for
  Materials and Surfaces (MatSurf), Turku, Finland}
  
\author{M. Ropo}

\affiliation{Department of Physics, Tampere University of Technology,
  PO Box 692, FI-33101 Tampere, Finland} \affiliation{COMP, Department
  of applied physics, Aalto University, Finland}

\author{B. Johansson}

\affiliation{Applied Materials Physics, Department of Materials
  Science and Engineering, Royal Institute of Technology, SE-10044
  Stockholm, Sweden}

\affiliation{Department of Physics and Materials Science, Uppsala
  University, SE-75121 Uppsala, Sweden}

\author{L. Vitos}

\affiliation{Applied Materials Physics, Department of Materials
  Science and Engineering, Royal Institute of Technology, SE-10044
  Stockholm, Sweden}

\affiliation{Department of Physics and Materials Science, Uppsala
  University, Box 516, SE-75120 Uppsala, Sweden}

\affiliation{Research Institute for Solid State Physics and Optics,
  Wigner Research Center for Physics, P.O. Box 49, H-1525 Budapest,
  Hungary}

\date{27 April 2015}

\begin{abstract} Iron-chromium alloys, the base components
    of various stainless steel grades, have numerous technologically
    and scientifically interesting properties. However, these features
    are not yet sufficiently understood to allow their full
    exploitation in technological applications. In this work, we
    investigate segregation, precipitation, and phase separation in
    Fe-Cr systems analysing the physical mechanisms behind the
    observed phenomena. To get a comprehensive picture of Fe-Cr alloys
    as a function of composition, temperature and time the present
    investigation combines Monte Carlo simulations using semiempirical
    interatomic potential, first-principles total energy calculations,
    and experimental spectroscopy. In order to obtain a general picture of the relation
    of the atomic interactions and properties of Fe-Cr alloys in bulk,
    surface, and interface regions several
    complementary methods has to be used. Using Exact Muffin-Tin Orbitals method the
    effective chemical potential as a function of Cr content (0--15
    at.\% Cr) is calculated for a surface, second atomic layer and
    bulk. At $\sim$10 at.\% Cr in the alloy the reversal of the
    driving force of a Cr atom to occupy either bulk or surface sites
    is obtained. The Cr containing surfaces are expected when the Cr
    content exceeds $\sim$10 at.\%. The second atomic layer forms
    about 0.3 eV barrier for the migration of Cr atoms between bulk
    and surface atomic layer. To get information on Fe-Cr in larger scales we use semiempirical methods. Using combined Monte Carlo molecular
    dynamics simulations, based on semiempirical potential, the
    precipitation of Cr into isolated pockets in bulk Fe-Cr and the
    upper limit of the solubility of Cr into Fe layers in Fe/Cr layer
    system is studied. The theoretical predictions are tested using spectroscopic measurements. Hard X-ray photoelectron spectroscopy and
    Auger electron spectroscopy investigations were carried out to
    explore Cr segregation and precipitation in Fe/Cr double layer and
    Fe$_{0.95}$Cr$_{0.05}$ and Fe$_{0.85}$Cr$_{0.15}$ alloys. Initial
    oxidation of Fe-Cr was investigated experimentally at 10$^{-8}$ Torr
    pressure of the spectrometers showing intense Cr$_2$O$_3$ signal. Cr segregation and the formation of Cr rich precipitates were traced by analysing the experimental spectral intensities with respect to annealing time, Cr content, and kinetic energy of the exited electron.
\end{abstract}


\maketitle

\section{Introduction}

Iron-chromium alloys have many technologically important and
scientifically interesting properties.\cite{vitos2011} On the other hand, growing technological challenges are faced in designing multifunctional steels. Developing of advanced steels to meet different standards simultaneously, e.g.\ high strength, proper workability and ductility, excellent corrosion resistivity, and specific magnetic properties all in various ambient conditions, require breakthrough innovations and cutting edge research.\cite{weng2011} 
Iron and
chromium are typical examples of a ferromagnet and an antiferromagnet,
respectively. As a function of composition and structure the magnetic
properties of Fe-Cr vary considerably, e.g. spin
glass\cite{burke.1983} and giant
magnetoresistance\cite{gruenberg.1986} features are found in these
systems. The crystallographic properties of Fe-Cr are also
peculiar. Although Fe-Cr has a body centered cubic (bcc) based
structure within the whole composition range, there exist both stable
and metastable composition regions in the phase
diagram. The experimental phase diagram of Fe-Cr at 300
  $^{\rm o}$C shows the miscibility gap beginning from 5--10 at.\% and extending to 90--95
  at.\% Cr. Within this miscibility gap two domains exist, regions of
  spinodal decomposition and nucleation and growth. The latter is
  located at the outskirts of the miscibility gap extending about 20
  at.\% on each side.\cite{danoix2000} On aging Fe-Cr often undergoes
transformation to either high-temperature $\sigma$-phase or separation
into Fe-rich ($\alpha$) and Cr-rich ($\alpha$')
phases.\cite{cieslak.2000,bonny2008,novy2009,pareige2011} 
  In 20 at.\% Cr alloy at 773~K temperature after 50 h annealing precipitates
  have been observed to occupy 2~\% of the alloy
  volume.\cite{novy2009} The microscale changes in the
crystallographic properties may induce considerable changes at
macroscopic level, e.g.\ the '475 $^{\rm o}$C
embrittlement'\cite{reidrich.1941,sahu.2009}, has significant effect
on the mechanical properties of certain steel grades. Fe-Cr is the
base component in many stainless steel grades due to the beneficial
properties of chromium. Certain amount of chromium makes an iron alloy
corrosion resistant\cite{Levesque2013}. At ambient conditions a thin
and transparent film of chromium oxide rapidly forms on the open
surface of the alloy preventing further oxidation and blocking
corrosion. The corrosion resistance of the ferritic stainless steels
increases abruptly by several orders of magnitude when the Cr content
in bulk reaches ${\sim}10$~at.\% level.\cite{Wranglen1985} This
 oxidation-related experimental threshold of Cr content
in bulk coincides with the calculated reversal point of the relative
magnitudes of the Fe and Cr chemical potentials in bulk and surface of
the Fe-Cr alloys. This reversal of the relative chemical
potentials enables the outburst of Cr on the otherwise
pure Fe surface found exclusively in the case of low-Cr Fe-Cr
alloys.\cite{ropo2007}  Therefore, at ambient conditions
  Cr$_2$O$_3$ is easily formed on the surface of Fe$_{\rm
    1-x}$Cr$_{\rm x}$ ($x \gtrsim 0.1$) alloys.  Furthermore, due to
  the strong tendency of Cr to segregate to the Fe-Cr/Cr$_2$O$_3$
  interface,\cite{Punkkinen2013} there is an additional driving force
  for a Cr$_2$O$_3$ layer to grow until the surface oxide reaches the
  protective nanometer scale thickness preventing the further
  oxidation of the material. Due to their technological importance and
  challenging open questions the surfaces of Fe-Cr alloys have
  recently received considerable scientific
  attention.\cite{geng2003,ackland2006,ponomareva2007,ropo2007,kiejna2008,
    ackland2009,Levesque2012,Levesque2013}

In this work, we investigate the physical conditions and possible
realizations of segregation, precipitation, and phase separation in
Fe-Cr systems. To get a comprehensive picture of the state of Fe-Cr as
a function of composition, temperature and time the present
investigation combines Monte Carlo simulations using semiempirical
interatomic potential, {\em ab initio} total energy calculations, and
experimental spectroscopy. Using several complementary methods it is
possible to get a more reliable picture of the interactions between Fe and
Cr atoms and explain the consequences of the atomic interactions for
the properties of Fe-Cr alloys. {\em Ab initio} methods are used to get the atomic scale energetics as a function of the concentration of the alloys, thermodynamics of large scale systems is obtained by Monte Carlo method, and experimental spectroscopy is used to probe the concentrations and atomic structure of real Fe-Cr systems as a function of Cr content and annealing time. However, one should carefully take into account the scope of validity of
each method. Keeping this in mind our aim is to give an extensive
picture of the atomic structure in bulk, surface, and interface
regions in Fe-Cr systems. The rest of the paper is divided into two
main sections and conclusions. The research methods are briefly
reviewed in Section II and the results are presented and discussed in
Section III.

\section{Methods}\label{method}

\subsection{First-priciples calculations}

The {\em ab initio} total energy calculations are based on the
density functional theory\cite{Hohenberg1964,Kohn1965} and were
performed using the Exact Muffin-Tin Orbitals (EMTO)
method.\cite{Vitos2001,Vitos2007} The basis set includes $s$, $p$, $d$,
and $f$ orbitals. The generalized gradient approximation in the PBE
form was used for the exchange-correlation
functional.\cite{Perdew1996} The total energy was calculated using the
full charge-density technique.\cite{Vitos2001a,Vitos2007} The alloys
were simulated as substitutionally disordered ferromagnetic bcc phase
using the coherent potential approximation (CPA) which provides the  continuous scanning of the concentration of the alloy.\cite{Soven1967} The
calculated equilibrium lattice constant was used for each
composition. The EMTO approach in combination with the CPA has been
applied successfully in the theoretical study of various structural
and electronic properties of alloys and compounds\cite{Vitos2007}
demonstrating the level of accuracy and efficiency needed also in the
present investigation. For more details of the electronic structure
calculations we refer to our earlier work. \cite{ropo2007,ropo2011}

The basic quantities used in the present study are surface energy
($E_{\rm surf}$), segregation energy ($E_{\rm segr}$), chemical
potential ($\mu$) and mixing enthalpy per atom ($E_{\rm mix}$). The
effective chemical potential ($\Delta\mu^{\rm b}$) and the slope of
the mixing enthalpy of bulk Fe-Cr are related within a simple
relation,\cite{ropo06}
\begin{equation}
  \Delta\mu^{\rm b}=(\mu_{\rm Fe}-\mu_{\rm Cr})^{\rm bulk} \approx 
  -\frac{\partial E_{\rm mix}}{\partial x} + {\rm constant}, \label{eqn1}  	
\end{equation}
where $x$ is the atomic fraction of Cr ($N_{\rm Cr}/(N_{\rm Fe}+N_{\rm
  Cr})$, $N_{\rm Fe}$ and $N_{\rm Cr}$ are the number of Fe and Cr
atoms in the investigated system, respectively). The surface energy is
defined as the energy needed to form a new surface per the formed new
surface area.
\begin{equation}
E_{\rm surf}=\frac{E_{\rm slab}-E_{\rm bulk}}{2A}, \label{eqn2} 
\end{equation}
where $E_{\rm slab}$ is the energy of the slab system with  two surfaces,
both having the area $A$. $E_{\rm bulk}$ is the energy of the bulk
system having the same amount of atoms as the slab system. 
The segregation energy of Cr from a region A to a region B is
defined as the energy needed to transfer a Cr atom from A to B and an
Fe atom from B to A.
\begin{equation}
E_{\rm segr}^{{\rm Cr: A}\rightarrow {\rm B}}= \Delta\mu^{\rm A}-\Delta\mu^{\rm B}.
\end{equation}

\subsection{Large-scale Monte Carlo simulations}

Because {\em ab initio} simulations for large systems and for longer
time scales are not possible we performed Monte Carlo simulations for
Fe$_{1-x}$Cr$_{x}$ alloys to test the {\em ab initio} predictions and
to investigate the {\em thermodynamic ground state} of the alloys in
larger scales.

Due to the time consuming simulations and large systems, {\em ab
  initio} methods can not be used. Thus the interatomic interaction
was modeled by a semiempirical potential, namely the two-band embedded
atom model (2BEAM) which is designed to reproduce the mixing enthalpy
of the Fe-Cr alloy.\cite{Bonny2011} One should note that the 2BEAM
potential is optimized mainly for bulk properties.

Precipitation and segregation of chromium in finite temperatures and
in large systems was studied using the Monte Carlo (MC) method where
possible moves included atom displacements and exchange of types (Fe
or Cr) of a pair of atoms. Displacements were performed with short
sequences of molecular dynamics (MD) simulations in the canonical
ensemble.  Using MD was observed to be more efficient in moving atoms
than the conventional Metropolis algorithm with atomic
displacements. All Monte Carlo--molecular dynamics (MCMD) simulations
were performed in NVT ensemble with the proper value of the lattice
constant obtained from separate NPT simulations as a function of
temperature and chromium concentration. It should be emphasized that
the MCMD calculations are pure equilibrium simulations; there is no
kinetics involved.

The MCMD method was used to study the near surface structure of the
Fe-Cr alloy and the structure of a iron--chromium interface in a
layered system. In the simulations a system of size $86\times 86\times
86$ \AA$^3$ with 54000 atoms was used. The lengths of the simulations
varied from 80000 to 120000 MC steps and results were calculated by
taking the averages of roughly 40000 last simulation steps.  For the
surface studies boundary conditions were applied in the $x$ and $y$
directions while leaving the two (001) $z$ surfaces open. In the case
of layer structure simulations periodic boundary conditions were
applied in all three directions. The interface orientation was (001).

Surface structure simulations were performed in temperatures of 300,
500, and 700~K and interface simulations in 300 and 700~K.

\subsection{Experiments}

The Fe/Cr bilayer was grown by electron beam physical
  vapour deposition from elemental Fe and Cr on a Si substrate. The top
  Fe layer of the Fe/Cr/Si sandwich was grown to about 50 nm thickness
  to protect the sample from contamination and mechanical failure. The
  thickness of the Fe film was checked by sputtering. The Fe-Cr alloy
  samples were prepared by induction melting under argon flow from
  elemental components.  The Cr bulk concentrations are 5 and 15 at.\
  \% for samples Fe$_{0.95}$Cr$_{0.05}$ and Fe$_{0.85}$Cr$_{0.15}$,
  respectively. The concentrations 5 and 15 at.\ \% Cr were selected
  to encompass the interesting concentration region of the onset of
  the corrosion resistance (9-13 at.\ \% Cr) in ferritic stainless
  steels and the ab initio prediction of the onset of the surface
  segregation of Cr in Fe-Cr alloys (8-9 at.\ \% Cr).\cite{ropo2007}
  Elemental components of purity better than 99.99 \% were used for
  all the samples. To start the investigations with fresh and
  unoxidized samples sputtering and annealing were used for all the
  samples. The Fe layers of Fe/Cr/Si were cleaned in the UHV of
  analyzer chamber of spectrometer by annealing at 150 $^{\rm o}$C
  followed by 20 min argon sputtering. To clean the alloy samples Ar
  sputtering was carried out until no traces of oxygen or carbon was
  detected. The atomic diffusion was driven by varying the temperature
  of the samples. The bilayer sample and alloys were heated to 500
  $^{\rm o}$C at rate 30 deg/min for different times and then cooled
  close to room temperature.

  The photoemission spectra were collected using both conventional
  x-ray photoelectron spectroscopy (XPS) (PHI ESCA 5400 Electron
  Spectrometer, Perkin Elmer) with non-monochromatic Al K$_\alpha$
  radiation at home laboratory and synchrotron radiation excited hard
  X-ray photoelectron spectroscopy (HAXPES) with high kinetic energy
  (HIKE) experimental station\cite{gorgoi2009} at KMC-1 beamline at
  Helmholtz-Zentrum Berlin (HZB), Bessy II.  VG Scienta R 4000
  electron analyzer, modified for electron kinetic energies up to 10
  keV and highresolution double-crystal monochromator were used. To
  obtain photon energies from 2300 to 7300 eV, Si(111), Si(311) and
  Si(422) crystals of the monocromator were selected for the presented
  measurements. The X-ray incidence angle was approximately 4$^{\rm
    o}$ in every experiment and the photoelectrons were detected in
  normal emission.  The energy scale was calibrated using Fermi level
  of the samples and Au 4f spectra of the calibration sample. The
  adjustable photon energy range from about 2 keV to 10 keV makes it
  possible to study photoelectrons with high kinetic energy which
  increases their inelastic mean free path (IMFP) making the HAXPES
  technique bulk sensitive as a comparison to surface sensitive
  laboratory XPS or soft x-ray range synchrotron radiation. Thus bulk
  sensitive investigations of atomic concentrations and chemical state
  of compound elements are possible without altering the original
  sample structure or chemistry by sputtering.  Also depth-profiling
  can be done by measuring core-levels with different binding energies
  and thus photoelectrons with different IMFPs or by exploiting
  different sampling depths of specific core electrons by adjusting
  the photon energy of radiation.  The high photon energy range makes
  it possible to measure photoelectrons with very high binding energy
  (low kinetic energy), for example Cr or Fe {\em 1s} at 5990 and 7110
  eV which enables also surface sensitive studies to be performed
  using HAXPES.

In addition we have used conventional Auger electron
  spectroscopy (AES) with Physical Electronics Model Cylindrical
  Mirror Analyzer to carry out depth profiling thorough the whole
  Fe/Cr bilayer before and after heating until Si substrate is
  reached. The depth profiles of the samples were obtained using
  Ar$^+$ sputtering (3000 V, $4\times 4$ mm$^2$ area, about 8 mPa
  Ar$^+$ pressure and 2 $\mu$A sample current). After every 1 minute
  sputtering cycle O KLL, Cr LMM and Fe LMM spectra from 400 to 760 eV
  were measured. The O and Cr spectra are overlapping and to get
  reliable quantitative information we used reference spectra of Fe,
  Cr and O (AlO) measured using same equipment and parameters. The AES
  profiles were obtained from differentiated spectra of Fe/Cr bilayers
  and reference samples after background subtraction using factor
  analysis delivered by Principal Component Analysis (PCA) of Casa XPS
  2.13.16 (Fig.\ \ref{fig:AES}).

\section{Results and discussion}\label{result}

We begin the structural analysis of Fe-Cr alloys by using the {\em ab
  initio} data to formulate some basic atomic models of the surface
segregation and bulk precipitation of Cr. Next step is to perform MC
simulations  for larger systems
to obtain results for more realistic cases. Finally,
we use photoelectron spectroscopy to experimentally determine the
segregation and precipitation in Fe/Cr double layer and Fe-Cr
  alloys as a function of annealing time and concentration.

\subsection{Initial relaxation in the surface region}
\begin{table}[ht]
\begin{center}
\begin{tabular}{l|cccccccc}
\hline
\hline
$c_{\rm b}$ (at.\%)& 0& 3& 5& 10& 12& 15& 20& 25\\\hline
$E_{\rm mix}$ (meV)& $0$& $-2.8$& $-1.8$& $10.2$& $ $& $27.7$& $44.3$& $60.0$\\
$E_{\rm segr}^{{\rm Cr: b}\rightarrow {\rm s}_1}$ (meV)& $251$& $ $& $ $& $-59$& $ $& $-16$& $60$& $ $\\
$E_{\rm segr}^{{\rm Cr: b}\rightarrow {\rm s}_1}$ (meV)& $216$& $204$& $145$& $-47$& $-71$& $-58$& $ $& $ $\\
$E_{\rm segr}^{{\rm Cr: b}\rightarrow {\rm s}_2}$ (meV)& $314$& $358$& $364$& $248$& $213$& $204$& $ $& $ $\\
\hline
\hline
\end{tabular}
\end{center}

\caption{Calculated (EMTO) bulk mixing enthalpy ($E_{\rm mix}$) and the 
    bulk to surface (surface layer: ${\rm s}_1$, second layer: ${\rm s}_2$) segregation energy of Cr ($E_{\rm segr}^{{\rm Cr: 
        b}\rightarrow {\rm s}_1,{\rm s}_2}$) of homogeneous Fe-Cr as a function 
    of atomic \% of Cr in bulk ($c_{\rm b}$).
    $c_{\rm b}= 3$ at.\% corresponds to the minimum of $E_{\rm mix}$. 
    Negative (positive) mixing enthalpy means stable (metastable) 
    bulk phase; negative (positive) $E_{\rm segr}^{{\rm Cr:
        b}\rightarrow 
      {\rm s}}$ means a driving force to a Cr containing (pure Fe) 
    surface. The two upper rows are calculated using 8 atomic layer slab in our previous work\cite{ropo2007} and the two lower rows are the present results of 12 atomic layer calculations.}\label{table:abinitio}
\end{table}

\begin{figure}[htb]
\includegraphics[width=0.40\textwidth,angle=0]{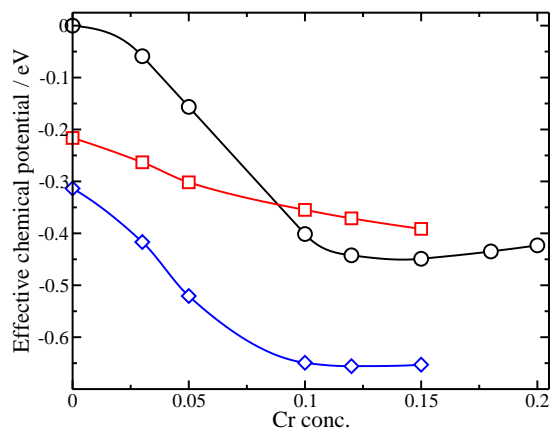}
\caption{(Color online) Calculated (EMTO) effective
    chemical potential $\Delta\mu=\mu_{\rm Fe}-\mu_{\rm Cr}$ of the
    surface atomic layer (red squares), second atomic layer (blue
    diamonds), and bulk (black circles) as a function of Cr content of
    the homogeneous alloy. The open symbols are calculated points, the
    curves are guides to the eye. In the present calculations we have
    used larger unit cells, 12 atomic (100) layers separated by vacuum
    of thickness equivalent to 6 atomic layers, compared to the
    previous work\cite{ropo2007} where 8 atomic layers of metal slab
    was separated by vacuum of thickness equivalent to 4 atomic
    layers.} \label{fig:chempot}
\end{figure}

Besides on external conditions, the structural evolution of an
initially homogeneous Fe-Cr alloy depends crucially on the
concentration of the alloy and the relative magnitudes of the atomic
diffusion rates at different spatial regions of the alloy. Using the
{\em ab initio} data shown in Table~\ref{table:abinitio} and in
Figure~\ref{fig:chempot} we can make predictions for the
structural evolution of initially homogeneous Fe$_{1-x}$Cr$_{x}$
alloys (same concentration throughout the whole bulk and surface
regions). Since atomic diffusion rates are usually significantly
higher in the near-surface regions than in the
bulk\cite{sladecek.2002} it is natural to split the consideration into
initial surface relaxation and more retarded bulk relaxation. For
non-homogeneous alloys, the structural evolution  depends also
on the gradients of the Cr concentration.

We begin our analysis by considering the relaxation of
the near-surface region connected to the bulk reservoir with fixed
concentration. According to the {\em ab initio} segregation energy
($E_{\rm segr}^{{\rm Cr: s}_2\rightarrow {\rm
      s}_1}=\Delta\mu^{s_2}-\Delta\mu^{s_1}$, s$_1$ and s$_2$ refer to
  the surface and second atomic layer, respectively,
  Fig.~\ref{fig:chempot}), the initial driving force ($E_{\rm
    segr}^{{\rm Cr: s}_2\rightarrow {\rm s}_1}<0$) within the two
  surface atomic layers pushes Cr atoms from the second layer to the
  surface layer. This driving force is increased by a factor of three, from $\sim98$ meV to
  $\sim295$ meV, when the Cr concentration of the alloy increases from
  0 at.\% to 10 at.\%. As Fig.~\ref{fig:chempot} shows, the driving
  force on Cr atoms between the bulk and the surface atomic layer is
  for low Cr alloys from the surface to the bulk, but when the Cr
  concentration exceeds $\sim10$ at.\% this driving force turns in the
  opposite direction pushing Cr atoms from the bulk to the
  surface. For all investigated alloy concentrations the second atomic
  layer forms a diffusion barrier for a Cr atom to move from the bulk
  to the surface atomic layer. The barrier starts from $\sim310$ meV at 0
  at.\% Cr and decreases to $\sim250$ meV at 10 at.\% Cr. This can be compared with other
  {\em ab initio} calculations\cite{Levesque2012} using 36 atom
  supercell including one Cr impurity atom and giving 355 meV barrier
  for the second atomic layer. From Fig. \ref{fig:eseg1} we see also
  that the model consisting of homogeneous bulk and two surface atomic
  layers describes the energetics of Fe-Cr (100) surfaces quite well
  because the segregation energy is practically converged to the bulk
  value at the third atomic layer.
  
To sum up, our {\em ab initio} calculations predict a Cr
  depleted surface atomic layer for Fe-Cr alloys below 10 at.\% Cr, a
  stable Cr enriched surface atomic layer within the bulk Cr content
  between 10 and 18 at.\% (a typical situation in many commercial
  steel grades), and a Cr containing stable surface beyond 18 at.\%
  Cr. The second atomic layer is predicted to be depleted in Cr. Since
  $E_{\rm segr}^{{\rm Cr: b,s}_2\rightarrow {\rm s}_1}$ increases with
  increasing Cr surface concentration, there exists a concentration
  dependent upper limit for the Cr content at the surface, as posed by
  the condition $E_{\rm segr}^{{\rm Cr: b,s}_2\rightarrow {\rm
      s}_1}=0$.\cite{ropo2007} 

\subsection{Bulk-surface relaxation}

Considering the thermodynamic ground state of the whole bulk-surface
Fe-Cr system, the bulk part should be relaxed too. The mixing enthalpy
of Fe$_{1-x}$Cr$_{x}$ (Table~\ref{table:abinitio}) suggests that if
the Cr concentration of the alloy exceeds ${\sim}10$~at.\%, the bulk
part of the alloy has a tendency to transform to an $\alpha$-$\alpha$'
phase separated system: Cr-rich precipitates immersed in the
Fe$_{0.97}$Cr$_{0.03}$ alloy. The Cr-rich $\alpha$' precipitates are
expected to avoid the contact with the surface because
the surface energy of Cr is higher than that of Fe. Therefore, a
low-Cr zone under the surface is expected to be formed
driving the  initially formed Cr-containing 
  surface back to the pure Fe surface. However, this can happen only
in vacuum. In ambient conditions the surface is expected
  to oxidize rapidly.  Because the Cr affinity to oxygen is much
  higher than that of Fe, Cr$_2$O$_3$ islands are expected to be
  formed on the surface. Due to the driving force of Cr to enrich to
  the metal/Cr$_2$O$_3$ interface \cite{Punkkinen2013} these islands
  can grow until the uniform protective oxide layer is formed on the
  surface. Therefore, Cr at the surface is bound to an
  oxide form, and the surface of Fe-Cr is practically in 
  an inert state during the retarded $\alpha$-$\alpha$' phase
separation in the bulk.

\subsection{Monte Carlo simulations}

\begin{figure}[htb]
\includegraphics[width=0.45\textwidth,angle=0]{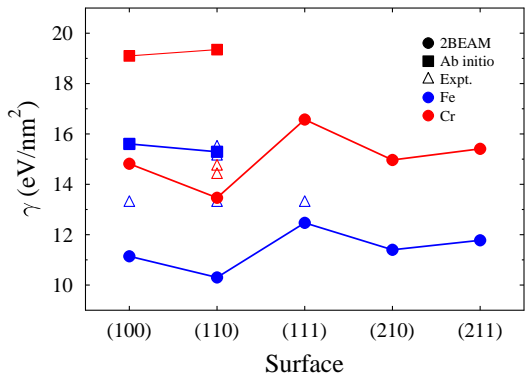}
\caption{(Color online) Surface energies of low index surfaces of bcc
  iron and chromium as predicted by the 2BEAM model. Results of {\em
    ab initio} calculations (filled squares) are from Ref.\
  \onlinecite{Punkkinen2011,Punkkinen2011prl}. Experimental values (open triangles) are
  from Ref.\ \onlinecite{Fu2009}.} \label{fig:esurf}
\end{figure}

\begin{figure}[htb]
\includegraphics[width=0.45\textwidth,angle=-0]{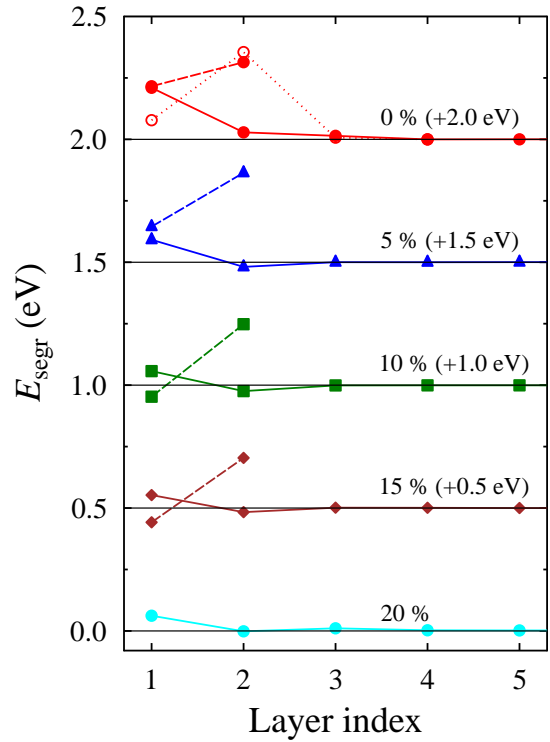}
\caption{(Color online) Segregation energy of a chromium atom in iron
  chromium alloy in the near surface atomic layers for different
  chromium concentrations as predicted by the 2BEAM model (solid
  lines) and {\em ab initio} (dashed lines) calculations (Table
  \ref{table:abinitio}). Curves are shifted by the marked amount for
  better visibility. The {\em ab initio} result for the pure iron from
  Ref.\ \onlinecite{Levesque2012} is plotted with dotted line and open
  symbols. } \label{fig:eseg1}
\end{figure}

\begin{figure}[htb]
\includegraphics[width=0.4\textwidth,angle=-0]{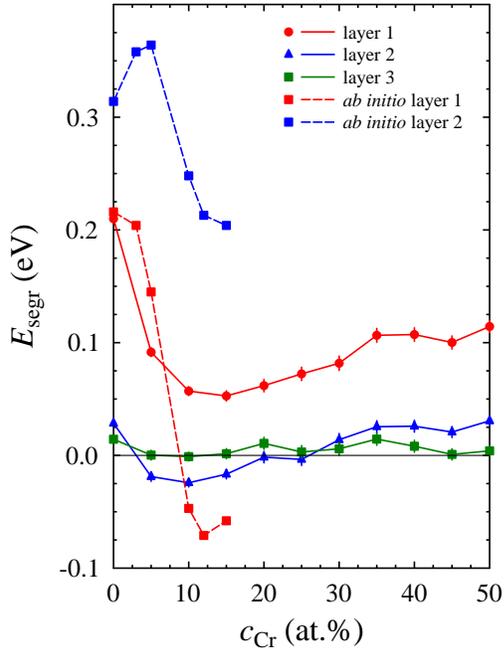}
\caption{(Color online) Segregation energy of the three near surface
  atomic layers as a function of bulk Cr concentration as predicted by
  the 2BEAM model. The {\em ab initio} data is the one presented in
  Table \ref{table:abinitio}.} \label{fig:eseg2}
\end{figure}

The predictions of the basic properties of Fe-Cr systems by the potential model are shown
in Figs.~\ref{fig:esurf}, \ref{fig:eseg1}, and \ref{fig:eseg2} where
the surface and segregation energies are shown. 

The energies of the (100) and (110) surfaces calculated by the 2BEAM
are considerably lower when compared with the {\em ab initio}
results. However, the differences between the energies of iron and
chromium are similar when comparing the 2BEAM and {\em ab initio}
results. Furthermore, all calculated energies differ from the
experimental values, which are also scattered and based partly on semiempirical estimates.

Surface segregation energies for different alloy concentrations were
calculated as averages of 1000 random alloy samples. Segregation
energy was defined as the energy difference between configurations
where the Cr atom was in the center of the simulation box and when the
atom was in one of the near surface atomic layers. One should note
that the variation of the segregation energy in different samples was
large (in the range 0.1--0.2 eV) compared to the energy itself. The
error bars (barely visible) in Figs.~\ref{fig:eseg1} and
\ref{fig:eseg2} are the errors of the mean. The figures show that
there is a barrier for chromium atoms to segregate to alloy
surface. However, for alloys of chromium concentrations
  in the range of 5--20 at.\% the segregation energy to the second
  atomic layer is negative predicting Cr segregation to the second
  layer.  Comparing the 2BEAM results with the {\em ab initio} data of
  Levesque et al.\cite{Levesque2012} at the Cr impurity level the
  2BEAM model predicts larger (smaller) segregation energy for the
  surface layer (second layer).  On the other hand, the EMTO
  segregation energy (Table \ref{table:abinitio}) is close to
  the 2BEAM value for the surface layer but close to the result of
  Levesque et al.\cite{Levesque2012} for the second layer.
Fig.~\ref{fig:eseg2} shows essentially the same data as
Fig.~\ref{fig:eseg1} but plotted along the Cr concentration axis. Here
the {\em ab initio} data is from Table
\ref{table:abinitio}. As Fig.~\ref{fig:eseg2} shows the
  EMTO results predict Cr segregation to the surface when Cr
  concentration exceeds 10 at.\%, whereas the MCMD results predict
  that the segregation of Cr to surface is prevented by an energy
  barrier but Cr segregation is expected to the second atomic layer
  within the range of 5-20 at.\% Cr concentration. One should also
note that there is a discrepancy between the two {\em ab initio}
results for surface layer of pure iron: 0.078 eV (Ref.\
\onlinecite{Levesque2012}) vs.\ 0.216 eV (Table
\ref{table:abinitio}). One notices from
  Figs.~\ref{fig:eseg1} and \ref{fig:eseg2} that for the segregation
  energy $E_{\rm segr}$ the {\em ab initio} methods predict stronger
  oscillations as a function of layer position or the concentration
  than the MCMD method. This can be partly related to the difference
  of model systems used in the calculations. In the {\em ab initio}
  calculations there are only one type surroundings for a Cr atom in a
  specific layer whereas in MCMD calculations, in principle, every Cr
  atom has a different surrounding. Therefore the MCMD results are
  'averaged' which possibly leads to reduced oscillations in $E_{\rm
    segr}$.

The difference between the results obtained by various
  computational approaches is expected to be mainly due to the
  specific approximations and implementations of the computational
  methods. The EMTO results were obtained relaxing the volume of the alloy uniformly (without local atomic relaxations) and random occupation of atomic sites was simulated using the CPA,
  i.e. the model system in our case is a homogeneous alloy.  Levesque et al.\cite{Levesque2012} used supercell approximation with fully relaxed atomic coordinates to model an impurity Cr atom. Their model system consisted of one Cr atom in the unit cell. The average overall bulk concentration in their calculations corresponds few at.\% Cr and the planar concentration of Cr impurity is 25 at.\% for (2a 2a 4a) slab ($a$ being the lattice constant). For layer relaxations of pure Fe they reported  $-0.02$ $\rm\AA$ for the surface layer and 0.04 $\rm\AA$ for the second layer.  A recent investigation based on the same {\em ab initio} method\cite{Punkkinen2011,Punkkinen2011prl} found $-0.002$ $\rm\AA$ for the surface and $0.04$ $\rm\AA$ for the second layer relaxations. This indicates that although there is a good agreement for the subsurface layer the top layer results are very different in these two pseudopotential {\em ab initio} investigations. Here we report similar behaviour for the segregation energies too. The differences between the results of Punkkinen et al.\cite{Punkkinen2011,Punkkinen2011prl} and Levesque et al.\cite{Levesque2012} could be partly due to the smaller slab and vacuum thicknesses and the using of constant-volume atomic relaxation in the latter investigation. The semiempirical potentials in the MCMD method are optimized mainly for bulk properties, which somewhat limits its feasibility for surface studies. 
  
  Since the accurate modeling of the surfaces and
  large bulk systems of Fe-Cr alloys at different temperatures is
  beyond the reach of any single computational technique we are forced
  to use several theoretical approaches to get the comprehensive overall
  picture. The results of Levesque et al.\cite{Levesque2012} are
  considered to describe the impurity in bulk, but due to the relatively small cell size these data cannot account for the impurity effects on the surface. Considering surface
  regions at higher Cr concentrations the EMTO method is expected to
  perform better and for large bulk systems, as a function of
  temperature, the MCMD is the natural choice.

\begin{figure}[htb]
\begin{center}
\includegraphics[width=0.5\textwidth,angle=0]{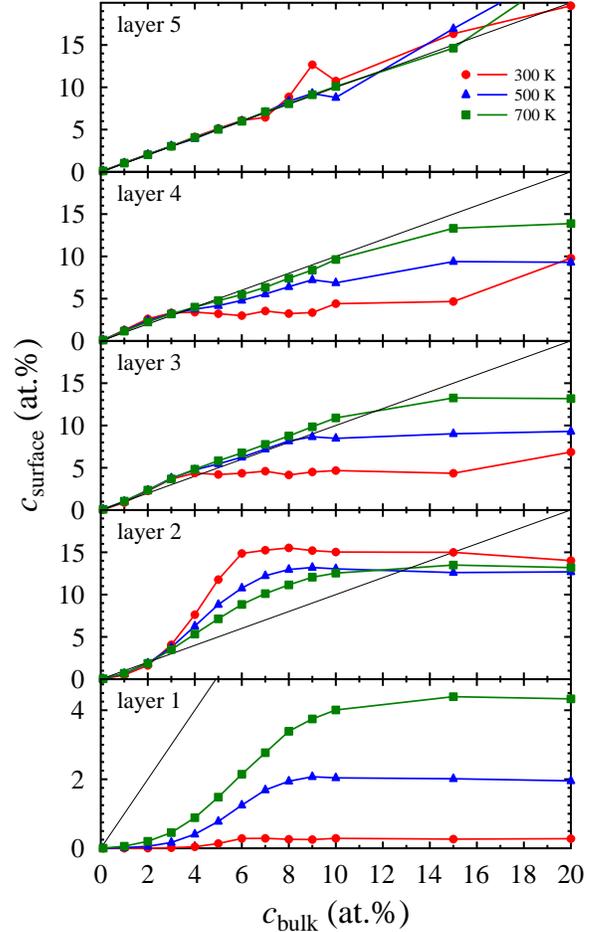}
\caption{(Color online) Chromium concentration of near-surface atomic
  layers as a function of the average bulk concentration.  Gray lines
  show the one-to-one relation between the bulk and layer
  concentrations. Note the different $y$ axis scale in the case of
  layer 1. } \label{fig:layerc_vs_bulkc2}
\end{center}
\end{figure}

Fig.~\ref{fig:layerc_vs_bulkc2} shows the near-surface concentrations
from the surface layer (layer 1) up to the fifth atomic layer (layer
5) as a function of bulk chromium concentration $c_{\rm bulk}$.  At
300~K the surface layer is exclusively occupied by Fe atoms up to
${\sim}5$ at.\% Cr in bulk. At that point the Cr concentration at the
surface jumps slightly upwards (to $\sim 0.3$ at.\%). With increasing
temperature the qualitative shape of the concentration curve of layer
1 does not change appreciably, but the height of the jump gets larger,
however, staying considerably lower than the average Cr concentration
in bulk (thin gray line in Fig.~\ref{fig:layerc_vs_bulkc2}). The MCMD
result for the bulk concentration threshold of the Cr containing
surfaces ($c_{\rm bulk}{\sim}$5--6 at.\%) compares reasonably well
with the {\em ab initio} results (8--9 at.\%).\cite{ropo2007} However,
the {\em ab initio} investigations predict the concentration of Cr at
the surface to exceed the bulk value whereas in MCMD simulations the
Cr concentration at the surface stays below the bulk value. This
is related to the fact that the 2BEAM model predicts a strongly
positive segregation energy for a Cr atom in the surface layer.

The plateau beyond ${\sim}10$ at.\% Cr in bulk seen in the
concentration curves of the layer 1 suggests that the Cr content in
nearby atomic layers have reached a certain saturation value.  This is
what actually happens as can be seen in the panel of layer 2: the Cr
content saturates to the ${\sim}13$ at.\% value at the bulk
concentration of $c_{\rm bulk}{\sim}8$~at.\%. With increasing depth
from the surface the layer resolved concentration curves gradually
approach the average bulk concentration line, as expected. However,
one should remember that the profiles shown in Fig.\
~\ref{fig:layerc_vs_bulkc2} are averaged concentrations parallel to
the surface plane direction. The bulk part of Fe-Cr is expected to be
$\alpha$-$\alpha$' phase separated at higher bulk concentrations,
where the thermodynamically optimal state consists of Cr-rich
precipitates ($c_\mathrm{Cr}=80-90$~at.\%) in a homogeneous Fe-Cr
alloy containing few per cents of Cr.  Since the surface energy of
Fe-Cr is in this model minimized by an Fe surface, the probability of
finding a Cr rich precipitation in the near-surface region is
low. This is also clearly seen in Fig.~\ref{fig:layerc_vs_bulkc2}. The
Cr concentrations of the third and fourth layers level
off to values below the average bulk value when the Cr content in bulk
is above the $\alpha$-$\alpha$' phase separation threshold (average Cr
content in bulk 6, 10, and 15 at.\%, for the 300, 500, and 700~K
simulations, respectively). In the case of layer 2 the increase of the
concentration above the average bulk value at $c_\mathrm{bulk}=3-13$\%
can also be attributed to the negative segregation energy of a Cr atom
in the second layer at these bulk concentrations.

\begin{figure}[htb]
\includegraphics[width=0.45\textwidth,angle=-0]{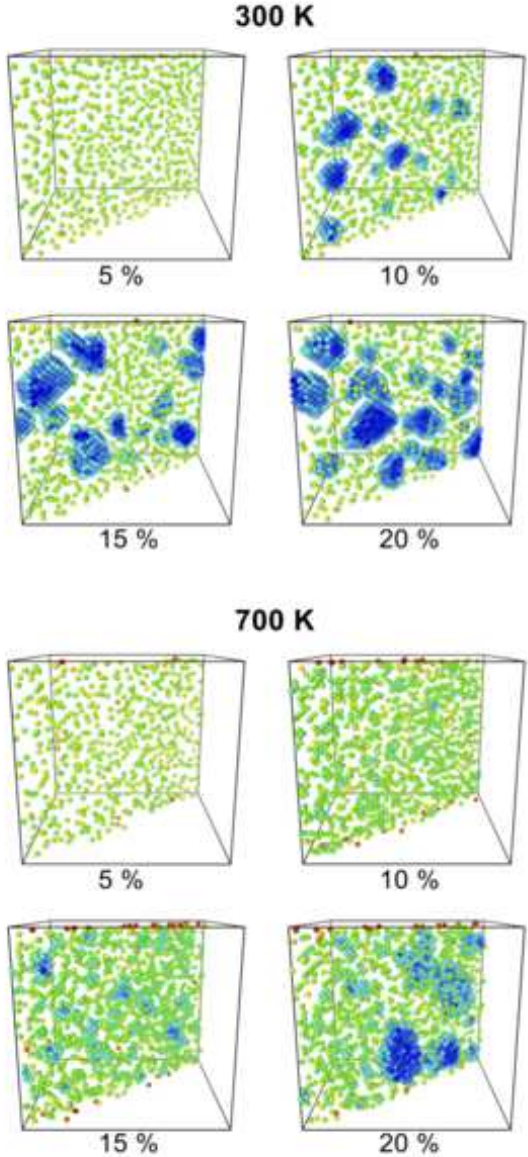}
\caption{(Color online) Snaphots of the simulated Fe$_{1-x}$Cr$_x$
  systems at different temperatures and with different chromium
  concentrations $x$. Figure shows slices of thickness 15 \AA\ in the
  (110) direction. Only chromium atoms are shown with color-coding
  according to potential energy (blue colour representing low potential energy). The top and bottom facets of the
  simulation box are open surfaces.} \label{fig:snapshots}
\end{figure}

\begin{figure}[htb]
  \includegraphics[width=0.5\textwidth,angle=-0]{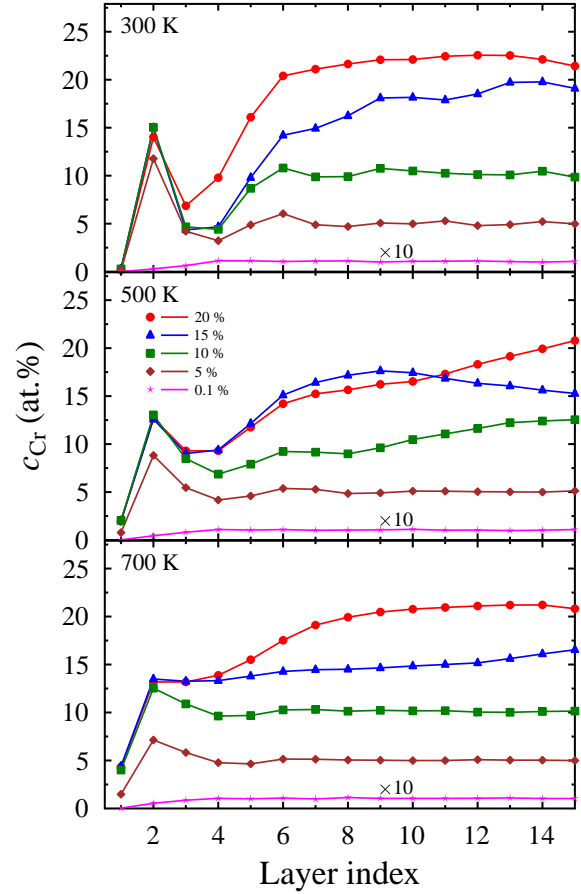}
\caption{(Color online) Chromium concentration depth profiles for
  different average chromium concentrations and simulation
  temperatures. Curves for 0.1 \% are multiplied by 10 for better
  visibility.} \label{fig:depthprofiles}
\end{figure}

In Fig.~\ref{fig:snapshots} the cross sections of the MCMD simulation
box are shown for concentrations $c_{\rm b}$ = 5, 10, 15, and 20 at.\%
at simulation temperatures 300~K and 700~K.  The low probability of Cr
rich $\alpha$' precipitates touching the surface is clearly
demonstrated. Adding Cr increases the number and size of Cr-rich
precipitates leaving the rest of the alloy in a homogeneous low-Cr
state in agreement with recent theoretical and
  experimental investigations.\cite{bonny2008,novy2009,pareige2011}
Fig.~\ref{fig:depthprofiles} shows the concentration profiles
corresponding to the MCMD simulations shown in
Fig.~\ref{fig:snapshots}. One can see that when the bulk concentration
$c_{\rm bulk}$ is below ${\sim}10$ at.\% the surface depth profile of
the Cr concentration shows only the short-period screening effect due
to the surface perturbation leveling off to the bulk concentration in
deeper layers. For concentrations $c_{\rm bulk} \gtrsim 10$~at.\%
long-period oscillations due to the Cr-rich precipitates show up
deeper in the sample.  An almost fixed ${\sim}13$~at.\% Cr value in
the second atomic layer is observed due to the interactions of Cr-rich
precipitates and the surface in line with the {\em ab initio}
predictions.

\begin{figure}[htb]
\includegraphics[width=0.5\textwidth,angle=-0]{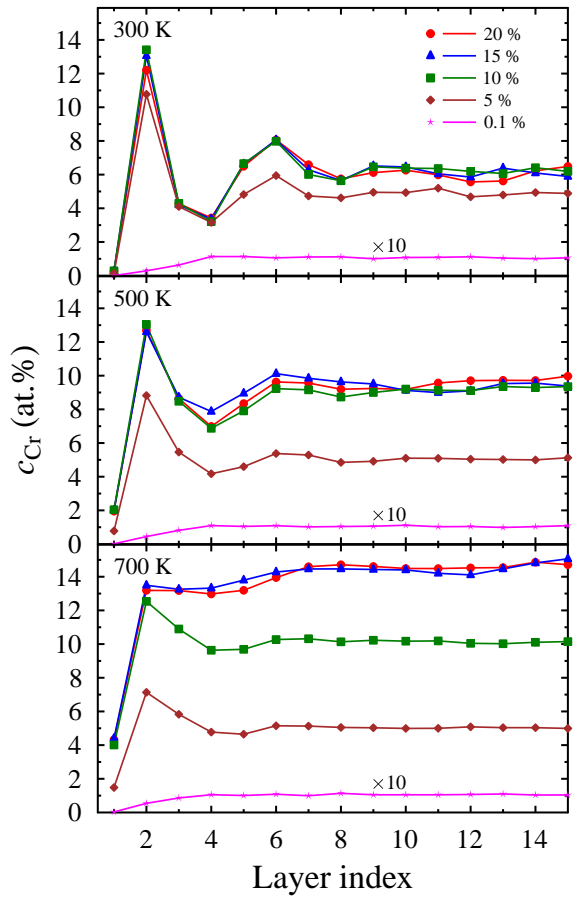}
\caption{(Color online) As in Fig.~\ref{fig:depthprofiles} but with
  chromium precipitates removed. Removing was done by only taking into
  account atoms whose average type during the last 60000 simulation
  steps was smaller than 1.5 where value 1.0 corresponds to pure Fe
  and 2.0 to pure Cr.} \label{fig:rp_depthprofiles}
\end{figure}

Fig.~\ref{fig:rp_depthprofiles} shows similar depth profiles as
Fig.~\ref{fig:depthprofiles} but with the Cr precipitates
removed. This was done by discarding all the atoms that had the
average atomic type index smaller than 1.5 where 1.0 corresponds to
pure Fe and 2.0 to pure Cr. We see that in the bulk part the
concentration reaches the solubility limit of the Cr in Fe (6, 10, and
15 at.\%, for the 300, 500, and 700~K simulations,
respectively). Concentrations of the surface layers are essentially
the same as in Fig.~\ref{fig:depthprofiles} because the chromium
precipitates do not extend to the surface layers. With the effect of
the precipitates removed, Fig.~\ref{fig:rp_depthprofiles} also shows
clearly the range and the decay rate of the concentration fluctuations
due to the screening of the surface perturbation. The low Cr content
in the first layer is overcompensated by the layer 2, which is back
compensated by the layers 3 and 4 etc. As expected, the compensation amplitude
decreases with increasing distance from the surface and the concentration fluctuations are less pronounced at higher temperatures.

\begin{figure}[htb]
\includegraphics[width=0.45\textwidth,angle=-0]{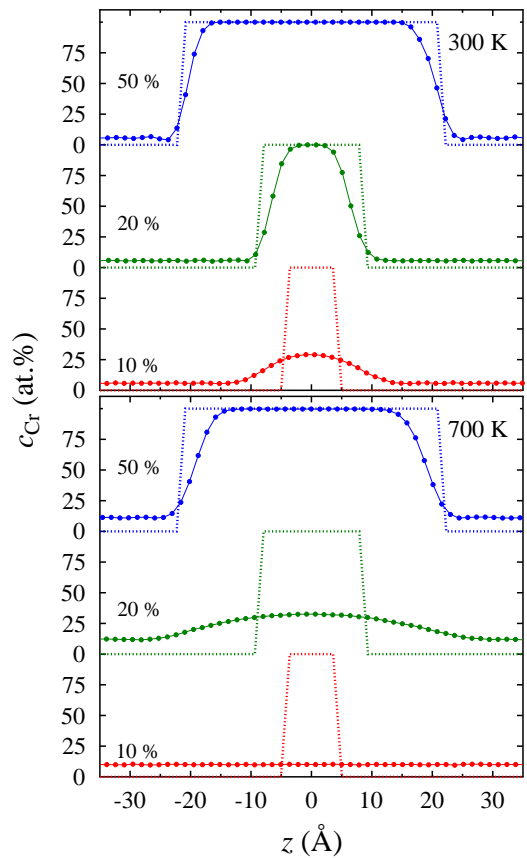}\\
\caption{(Color online) Cr concentration profiles for the layer widths
  corresponding different total Cr concentrations simulated in
  different temperatures. Thin dashed lines show the initial Cr
  profile.} \label{fig:layerconc}
\end{figure}

Considering the experimental investigations of segregation
  and precipitation in Fe-Cr it would be beneficial to study, in
  addition to alloys, also Fe/Cr layer structure. In the layer
  structure one can study the evolution of Fe-Cr as a function of Cr
  concentration in a more transparent way because the density gradient is a one-dimensional function. Similarly in this case the
  segregation of Cr in the surface of the Fe layer can also be
  investigated in a controlled way. These benefits in mind, we finish
  our simulations by considering Fe/Cr layer structure.
Fig.~\ref{fig:layerconc} shows the results of the simulations of the
Fe/Cr interface. The dashed curves are the initial states in which
there exist pure Cr and Fe sections in the simulation cell, whereas
the solid lines show the situation after the simulation. The results
show that the thermodynamical ground state of the investigated system
consists of two different parts: low Cr section where the average Cr
concentration is $\sim 6$ at.\% at 300~K ($\sim 12$ at.\% at
700~K, which compares well with experimental results of
  about 14 at.\% at 773 K\cite{novy2009}) and the excess of the Cr
remains in the original Cr part.  In the case of temperature of 700~K
and original concentration of 10~at.\% the whole Cr layer is dissolved
into the Fe. This is expected as the solubility of Cr in Fe at this
temperature is over 10 at.\%. At the lower temperature of 300~K the
original layer is partly dissolved leaving precipitates with mainly
(110) facets. Also, for the 700~K case with 20 at.\% chromium the
layer structure changes into a single precipitate again with (110)
facets.  The appearance of (110) facets is in line with {\em ab
  initio} interface calculations\cite{lu2011} which show that (110)
interface has the lowest energy which compares well with the surface
energies in Fig.~\ref{fig:esurf}.  The 300~K, 50~at.\% case shows
oscillations of the concentration profile which are caused by the
similar compensation effect as on the surface and partly by the
faceting of the interface.  These oscillations are washed out when
temperature is raised to 700~K and the positions of the atoms become
more random.

\subsection{Segregation, precipitation and oxidation in Fe-Cr systems}
\subsubsection{Cr segregation in Fe/Cr bilayer and Fe-Cr alloys}
\begin{figure}[htb]
\includegraphics[width=0.485\textwidth,angle=-0]{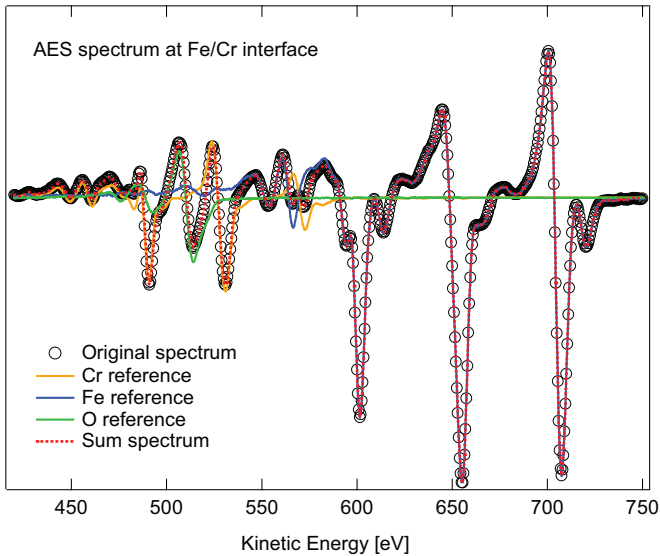}\\
\caption{(Color online) Example of an AES spectrum of Fe/Cr
  interface (o curve) of Fe/Cr/Si sample. The sum spectrum (dashed
  curve) was obtained by Principal Component Analysis (PCA) of Casa
  XPS 2.13.16.\cite{CASA_XPS} \label{fig:AES}}
\end{figure}

Experimentally the segregation of Cr can be demonstrated by comparing
the concentrations of Cr and Fe atoms as a function of the probing
depth. We have used two different methods to carry out depth
profiling. In AES the topmost atoms are removed by ion gun layer by
layer and after every sputtering cycle the Cr and Fe Auger electron
spectra (Fig.\ \ref{fig:AES}) are collected and the intensities of the
compound elements are compared. However, due to ion bombardment
sputtering is considered as a destructive method and it may have
unwanted effects on some concentrations values. For example,
preferential sputtering or reduction of oxides can be a drawback in
some studies. Preferential sputtering is not a problem in case of Fe
and Cr but the analysis of oxide layer can be.  This technique is not
ideal for chemical analysis which is why HAXPES spectra measured using
different photon energies (different IMFP of photoelectrons) was
important to collect as well.  Due to the limited time for synchrotron
radiation experiments it is impossible to measure HAXPES spectra layer
by layer but use of three selected photon energies and analysis of
different core-level spectra bring us information on the
concentrations of Fe and Cr at different depths from the surface.  The
investigated Fe/Cr double layer and Fe-Cr alloys are known to have
polycrystalline structure containing grain boundaries and compound
intermixing and diffusion around Fe/Cr interface and oxygen diffusion
into the layers \cite{Fedchenko2012} which can complicate the
investigation of Cr segregation and interpretation of the results. AES
together with XPS was used to carry out preliminary diffusion and
depth profile experiments.
\begin{figure}[htb]
\includegraphics[width=0.49\textwidth,angle=-0]{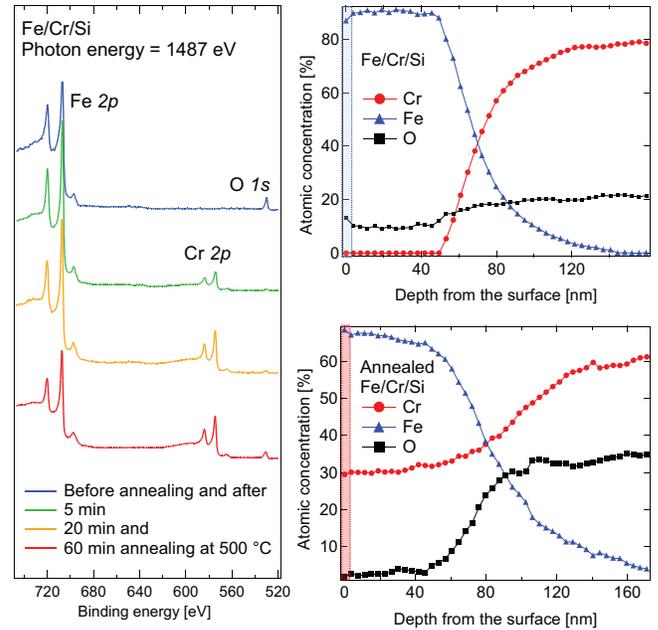}\\
\caption{(Color online) Left hand side presents the Fe 2p,
  Cr 2p and O 1s XPS spectra of a Fe/Cr bilayer sample before (topmost
  spectrum) and after annealing. These spectra were measured using Al
  K$\alpha$ twin anode. Photoemission spectra clearly demonstrate the
  increase of Cr 2p intensity as a function of annealing time
  confirming the Cr diffusion into the Fe layer.  The upper AES
  profile on the right is taken over the Fe/Cr interface before any
  heating was carried out starting sputtering from the surface
  represented by upper most XPS spectrum (blue) on the left. The lower
  AES depth profile on the right was taken after all the heating
  treatments starting from the surface concentration presented in the
  lowest (red) XPS spectrum on the left.} \label{fig:XPS_AES}
\end{figure} 
AES depth profiles that were measured from
  untreated and heated Fe/Cr/Si layer samples are presented in Fig.\
  \ref{fig:XPS_AES}. Annealing was performed in XPS located in the
  same room than AES but missing the possibility to transfer the
  sample in UHV between the spectrometers. XPS spectra were measured
  before and after every heating period (Fig.\ \ref{fig:XPS_AES})
  which were carried out at 500 $^{\rm o}$C for 5, 15 and 40 minutes
  giving total heating time of 60 minutes. The pressure of the
  preparation chamber where the heating was performed was $3\times
  10^{-8}$ Torr at maximum. Sample was cooled close to room
  temperature before measuring the spectra. Annealed sample was
  transferred to AES through air and two minutes of sputtering was
  performed before starting the depth profiling to get rid of the
  carbon contamination. The upper depth profile in Fig.\
  \ref{fig:XPS_AES} nicely presents the well separated Fe and Cr
  layers. It must be noted that the signal in the topmost XPS spectra
  at the same binding energy as Cr 2s would appear (around 696 eV)
  is entirely due to the non-monochromatic radiation used in
  laboratory XPS. About 10 \% oxygen concentration in the Fe layer is
  comparable to the topmost XPS spectrum in Fig.\
  \ref{fig:XPS_AES}. The content of oxygen through the whole sample is
  quite high but can be explained by the fact that polycrystalline
  samples contain number of grain boundaries or dislocations which can
  enhance the oxygen diffusion or diffusional transport of Cr towards
  the surface.\cite{Ostwald2004} Also the pressure during the sample
  growth and sputtering with Ar$^+$ ions can have effect on the oxygen
  concentration within the layers. The oxygen content increases in
  phase with Cr content in both profiles which in addition to the
  above mentioned reasons is related to higher affinity of Cr for
  oxygen and Cr reacting with inward diffused
  oxygen.\cite{Camra2005,Pujilaksono2011, Horita2008} After heating
  the Fe/Cr bilayer for 60 min at 500 $^{\rm o}$C the layered
  structure has been destroyed and the Cr concentration within the
  first 40 nm is about 30 \% increasing after that up to 60 \% before
  any signal from Si substrate was observed. The oxygen concentration
  for the annealed sample is less than 5 \% until fast decrease in Fe
  and slow increase of Cr concentration takes place at approximately
  same depth from the surface where the original Fe/Cr interface was
  detected before heating. Oxygen concentration increases about the
  same percentage value as Cr concentration but more rapidly. This
  is expected since the diffusivity of O in bcc Fe\cite{Shang2014} is
  much higher than diffusivity of Cr in Fe. \cite{Takasawa2002} Most
  probably the reason for low oxygen concentration within the first 40
  nm in annealed double layer sample is due to the formation of
  protective passive layer on the surface because of increased Cr
  concentration. The concentration at the uppermost atom layers of
  annealed sample is not visible in Fig.\ \ref{fig:XPS_AES} since some
  sputtering was done prior to the presented data to remove the
  contamination (carbon) that was absorbed on the surface during the
  sample transfer from XPS to AES.

\begin{figure}[htb]
\includegraphics[width=0.485\textwidth,angle=-0]{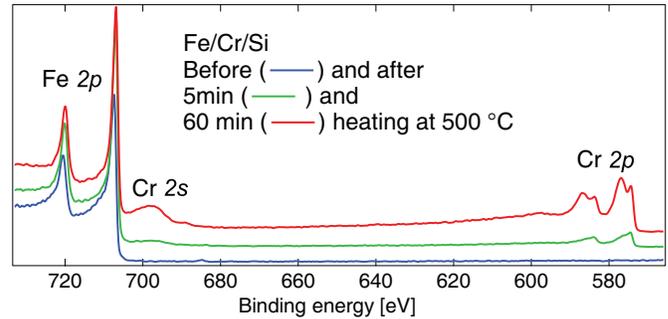}\\
\caption{(Color online) Overview HAXPES spectra of the
  Fe/Cr bilayer sample measured with 2300 eV photon energy before and
  after two heating treatments. In addition to the Cr 2s and Cr 2p
  spectra appearing after annealing, the increased intensity of Cr 2p
  as a comparison to Fe 2p after different annealing times
  demonstrates the Cr diffusion towards the surface. The intensity of
  Fe 2p3/2 of the presented spectra is normalized to one after setting
  background to zero.} \label{fig:Fig12} 
\end{figure}

\begin{figure}[htb]
\includegraphics[width=0.50\textwidth,angle=-0]{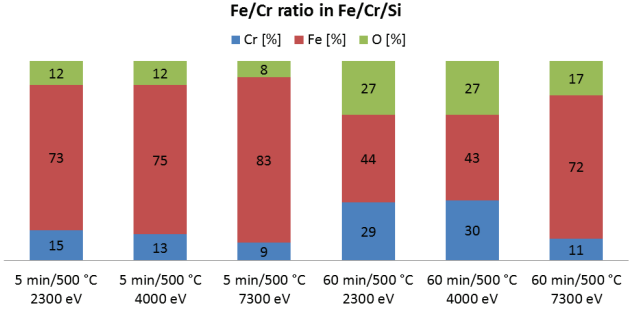}\\
\caption{(Color online) Concentrations of Cr, Fe and O in
  Fe/Cr/Si. Relative intensities are calculated by analyzing the area
  of Cr 2p, Fe 2p and O 1s spectra.} \label{fig:Fig13}
\end{figure}

To investigate the Cr segregation and also precipitation in Fe-Cr
systems and to test our theoretical predictions HAXPES technique was
used to monitor the concentration profiles of the Fe/Cr double layer
and chemical state of Fe and Cr atoms in two Fe-Cr alloys as a
function of photon energy and heating time. Investigations were
performed with photon energies from 2300 to 7300 eV in order to study
the concentration profiles of the chemical components as a function of
the probing depth. Prior to HAXPES experiments several annealing tests
were performed with laboratory XPS to find the right annealing
temperatures and time. The ideal temperature for Cr segregation in
Fe/Cr/Si turned out to be 500 $^{\rm o}$C. At this temperature Cr
diffused towards the surface in a few minutes time so that we were
able to detect it with XPS. In the proper HAXPES segregation
experiments the bilayer sample was heated in several steps as
described above to gradually follow the Cr diffusion into the Fe layer 
(Fig.\ \ref{fig:Fig12}).
Before starting the HAXPES segregation study the Fe layers of Fe/Cr/Si
were cleaned in the UHV of analyzer chamber of HIKE spectrometer by
annealing at 150 $^{\rm o}$C followed by 20 min Ar$^+$ sputtering. To
start the Cr segregation investigations the quality of the Fe/Cr
double layer structure, i.e. the absence of Cr in the Fe layer, was
checked by scanning the binding energy range of Cr 2p and Cr 1s core
levels before any heating was performed. Initially no Cr was found
using 2300 eV, 4000 eV and 7300 eV photon energies i.e. experimental
conditions that correspond to IMFP ($\lambda$) values from 2 to 7 nm
and thus sampling depths (3$\lambda$) of approximately 9-20 nm
depending on the oxidation and core-level
studied. \cite{Tanuma2003,Olsson2011}  It is worth mentioning that no
Si was found in the spectra, either. This indicates the proper double
layer structure of the sample, complete Fe layer on top of a Cr
layer. Due to sputtering the thickness of Fe layer before any heating
treatments was less than in the AES studies being still more than 20
nm. After the cleaning procedure sample was heated to 500 $^{\rm o}$C
sequentially for 5, 15 and 40 minutes, total heating time being 5, 20
and 60 minutes, respectively. The spectra were measured after each
heating period when the sample was cooled close to room
temperature. The gradual changes of the atomic concentration profiles
were observed with heating time (Fig.\ \ref{fig:Fig12}). These spectra
nicely demonstrate the appearance of Cr spectra for heated samples.
Before heating no signal at Cr 2p binding energy was observed but
already after 5 min heating Cr 2s and Cr 2p signals appeared as shown
in Fig.\ \ref{fig:Fig12}. The cross sections or spectrometer
transmission are not taken into account in Fig.\
\ref{fig:Fig12}. which slightly underestimates the real intensity of
Cr 2p and 2s photoemission signals relative to that of Fe 2p. The bulk
concentrations of O, Cr and Fe after short and longer heat treatments
are shown in Fig.\ \ref{fig:Fig13}. Intensities presented in Fig.\
\ref{fig:Fig13} are calculated by comparing the area of O 1s, Fe 2p
and Cr 2p spectra where Scofield cross-sections,\cite{scofield1973}
spectrometer transmission and other experimental parameters have been
taken into account. The Cr concentrations presented in Fig.\
\ref{fig:Fig13} are in line with the AES depth profile in Fig.\
\ref{fig:XPS_AES}. The oxygen concentration for heated sample derived
from the HAXPES spectra is higher than the AES profiles give but this
is most probably due to the pressure difference during the heating and
measurements. Measurement with 2300 eV photon energy gives the average
Cr concentration within 5 or 6 nm thick layer to be close to 15 at.\
\% (Fig.\ \ref{fig:Fig13}). The concentrations in the 4000 eV photon
energy measurements are similar but when compared to the more bulk
sensitive, 7300 eV photon energy case, the Cr concentration is much
lower than the Fe concentration. The increasing Cr concentration in
the direction from bulk to the surface within the Fe layer of the
heated sample is in line with our earlier
calculations.\cite{ropo2007,Punkkinen2013} In Fe$_{0.95}$Cr$_{0.05}$
and Fe$_{0.85}$Cr$_{0.15}$ alloys the Cr segregation was not so
evident. In bulk (comparison of Fe 2p and Cr 2p spectra) almost no
change in Fe/Cr intensity ratio was detected but more surface
sensitive measurements of Cr and Fe 1s photoelectrons show some
increase in Cr intensity as a result of heating, especially in the
oxidized Cr. 

\subsubsection{Oxidation}
\begin{figure}[htb]
\includegraphics[width=0.40\textwidth,angle=-0]{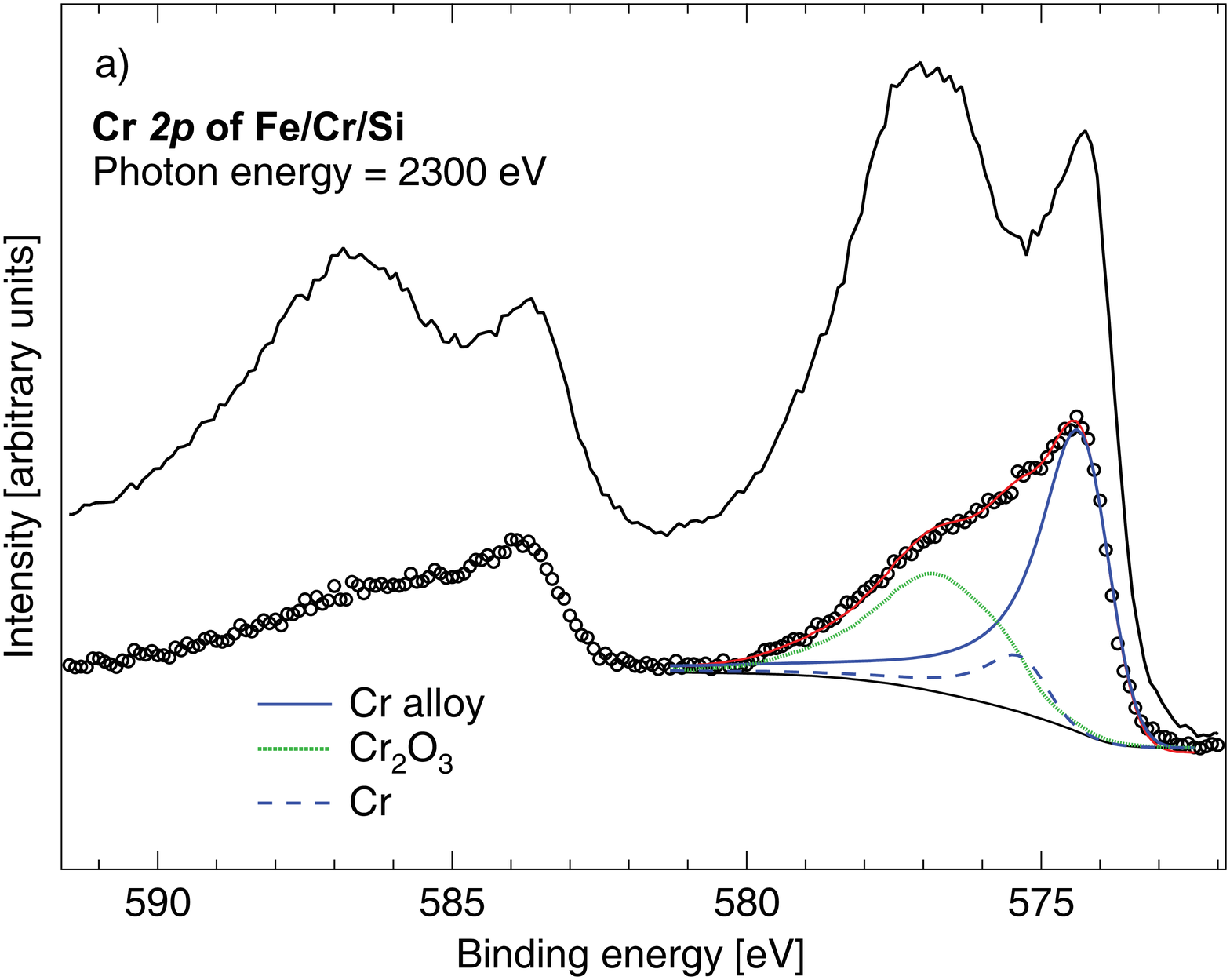}
\label{fig:Fig14a}
\vspace{2mm}
\includegraphics[width=0.40\textwidth,angle=-0]{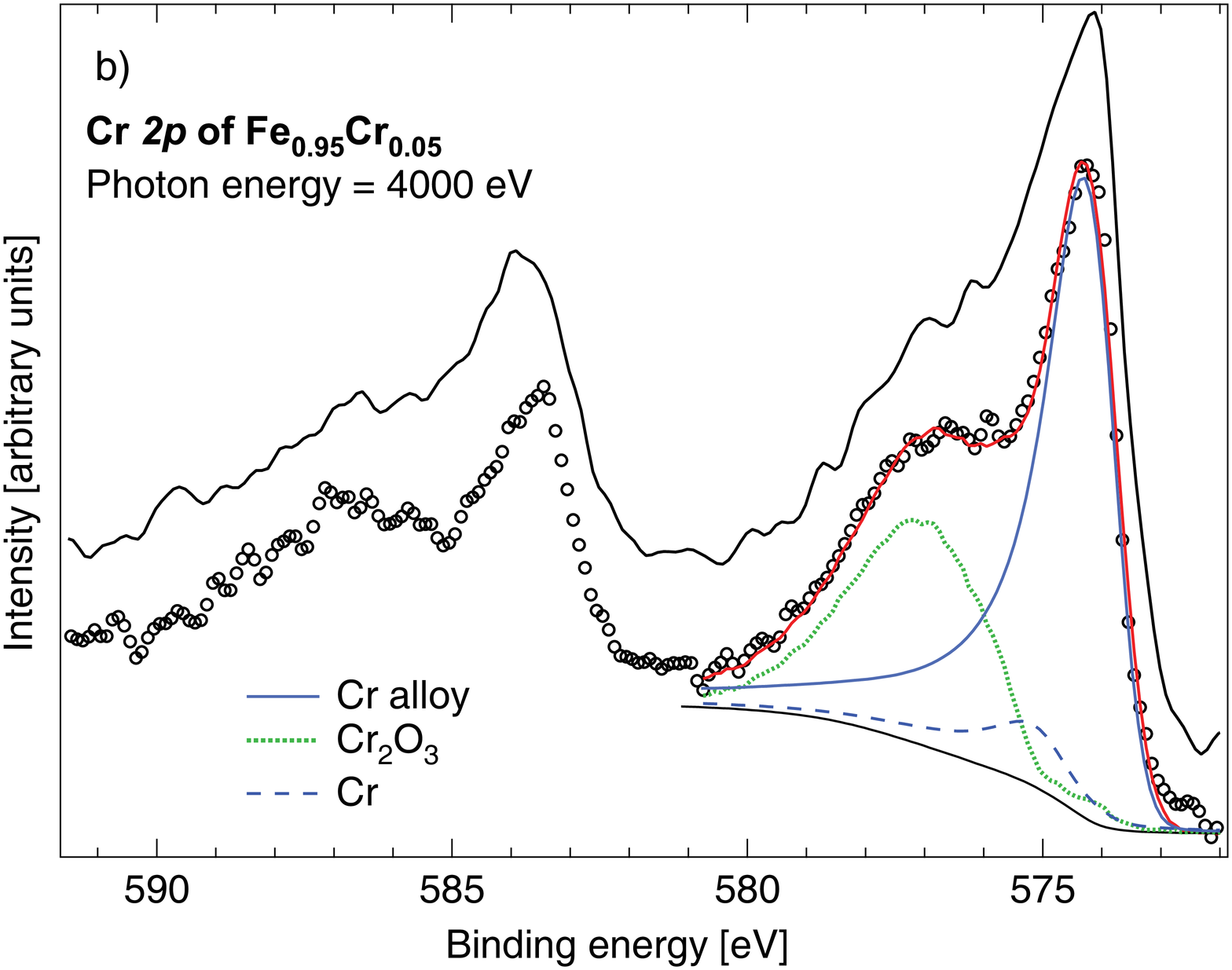}
\label{fig:Fig14b}
\vspace{2mm}
\includegraphics[width=0.40\textwidth,angle=-0]{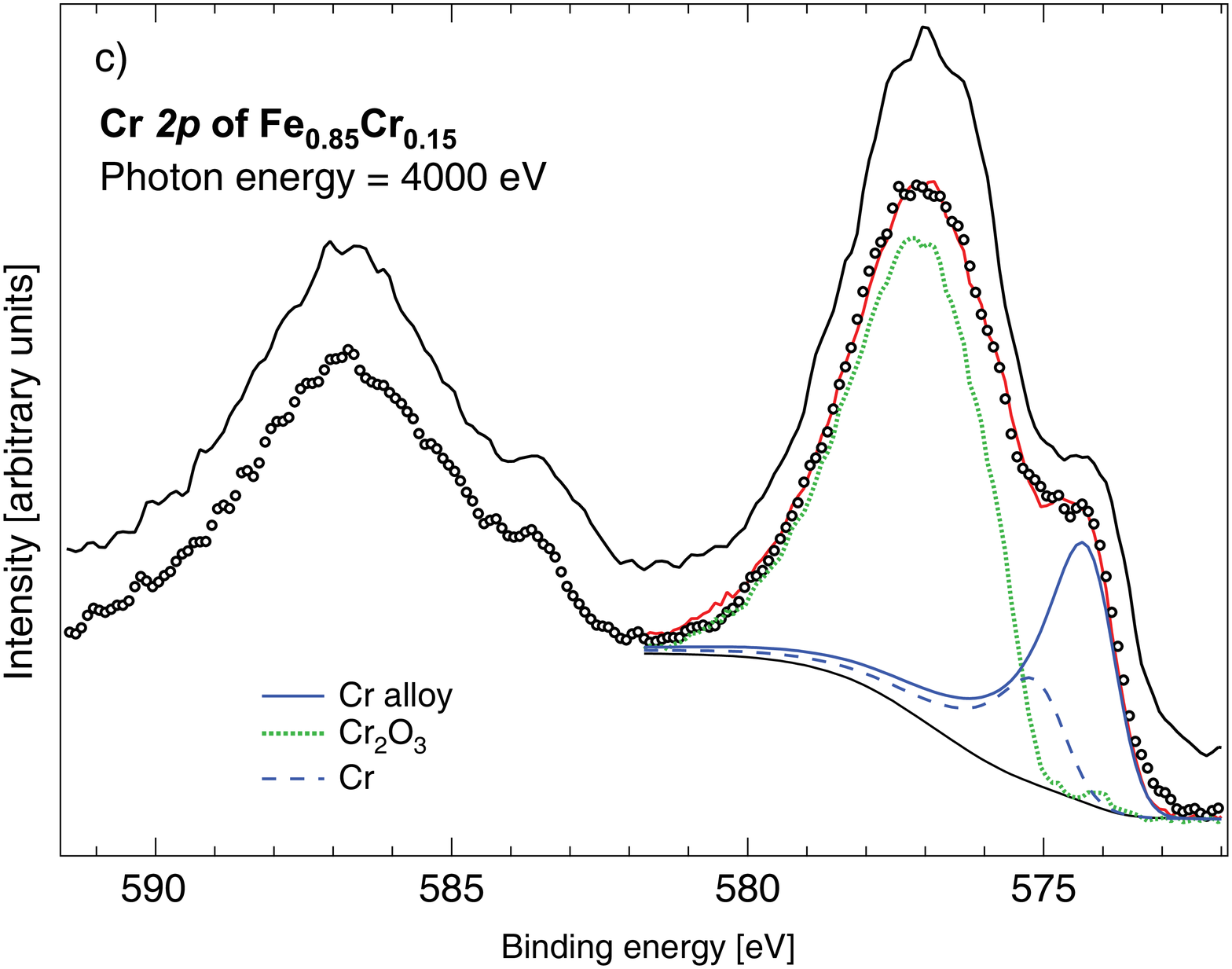}
\label{fig:Fig14c}
\caption{(Color online) Cr 2p spectra of Fe/Cr bilayer
  sample after 5 (open circles) and 60 min (solid
    line) heating (a) and Cr 2p spectra of Fe$_{0.95}$Cr$_{0.05}$ (b) and
  Fe$_{0.85}$Cr$_{0.15}$ (c) alloys before and after heating 10 minutes at
  500 $^{\rm o}$C. The lower spectra in a, b
  and c are fitted using three components to represent signals from
  bulk like Cr, Cr from Fe-Cr alloy and oxidized
  Cr.} \label{fig:Fig14}
\end{figure}

Even after 5 min annealing at 500 $^{\rm o}$C at pressure of about
$5·\times 10^{-8}$ Torr the Fe/Cr bilayer sample and Fe-Cr alloys were
evidently oxidized enabling  the  investigation  of  the  ratio  of
alloy Cr and oxidized Cr. Oxidized Cr is well resolved on the high
binding energy side of the alloy Cr (Fig.\ \ref{fig:Fig14}). A higher
amount of oxidized Cr in Fe/Cr bilayer was observed in the more
surface sensitive (photon energy=2300 eV) Cr 2p HAXPES experiment
(Fig.\ \ref{fig:Fig14}a) compared to the 4000 eV or 7300 eV
measurements \cite{Kokko2013} indicating increasing Cr concentration
towards the open surface also in the initial oxidation
experiments. The spectra show also that, especially at room
temperature range, oxygen penetration deeper into the metal is slow. 

Fig.\ \ref{fig:Fig14} presents Cr 2p photoemission spectra of bilayer
sample after 5 and 60 minutes heating at 500 $^{\rm o}$C measured with
2300 eV photon energy (a) together with Cr 2p spectra of
Fe$_{0.95}$Cr$_{0.05}$ and Fe$_{0.85}$Cr$_{0.15}$ alloys before and
after heating 10 minutes at 500 $^{\rm o}$C (b and c, respectively).
The spectra that were measured after 5 min annealing of Fe/Cr/Si (a)
and before annealing of alloys (b, c) present a fit with three
components. The two components, Cr and Cr alloy can be resolved due to the chemical shift between Cr atom surrounded mostly by other Cr atoms or Fe atoms. Thus segregation of Cr and formation of Cr dominating areas can be estimated by analyzing the fitted spectra. The broad feature from approximately 575 to 580 eV in all the spectra is the spectrum of measured Cr$_2$O$_3$ reference sample detected with same parameters than the actual samples. The appearance of this pronounced feature on the high binding energy side of the bulk and alloy Cr makes it possible to distinguish the oxidized Cr from the unoxidized one.

In Cr oxides the interaction between valence band 3d electrons and 2p vacancy core levels can create a number of final states which is called multiplet splitting \cite{biesinger2011,Gupta1975}. For example, Biesinger et al. \cite{biesinger2011} have presented a fit for asymmetric Cr 2p photoemission spectra of Cr(III) oxide that consists of five components. In this study we have used Cr$_2$O$_3$ reference spectrum to replace the multiple feature fit that otherwise would have been needed to properly describe the complicated multiplet splitting of Cr(III) oxide\cite{biesinger2011}. However, the Cr 2p photoemission spectra can't be fitted satisfactorily by using only Cr$_2$O$_3$ reference spectrum and one asymmetric Voigt-line shaped feature to describe the unoxidized Fe-Cr alloy. An additional component is needed between Cr$_2$O$_3$ and Cr alloy features. This structure at approximately 574.4 eV binding energy originates from bulk like chromium (Cr-Cr) (labelled Cr in Fig.\ \ref{fig:Fig14}). The shift between the Cr alloy and Cr components in the fitted spectra is approximately $0.9-1$ eV, which is close to the value we observed in our HAXPES measurements of pure, unoxidized Cr and Cr$_2$O$_3$ reference samples. The asymmetry parameter and full width at half maximum (FWHM) value used for Cr and Cr alloy line shape in the fitting procedure were derived from the HAXPES spectra of Cr bulk reference. In addition to the experimental arguments first principles calculations within density functional theory using the complete screening picture \cite{Olovsson2010} were performed to estimate the binding energy shift between bulk Cr and Fe$_{1-x}$Cr$_x$ alloys of different concentrations. These calculations gave a negative binding energy shift for Cr 2p core level spectra which was between $-0.1$ and $-0.5$ eV for Cr concentrations $x=0.1-0.9$ being largest in case of  Fe$_{0.8}$Cr$_{0.2}$. 

Even though the samples
were measured in ultrahigh vacuum the Cr oxidation was
evident. However, almost no sign of Fe oxide was seen in the Fe
photoemission spectra. As presented already in Fig.\ \ref{fig:Fig14}
intensity of Cr in Fe/Cr/Si further increased between the 5 and 60
minutes heating and in the latter spectrum larger number of Cr atoms
is detected as an oxide than in metallic form (Fig.\
\ref{fig:Fig14}a). During  the  60  min  annealing  the  amount  of
Fe  decreases (Fig.\ \ref{fig:Fig13}) to lower level due to Cr
segregation (Fig.\ \ref{fig:Fig14}a) and the oxidized form of Fe is
not resolved even on the surface. As a comparison to unoxidized
(sputtered until no traces of O or C were detected) bulk Fe reference
sample only the surface sensitive Fe 1s spectrum of Fe/Cr/Si (not shown
here) that was measured before any heating shows slightly increased
intensity on the high binding energy side where oxidized Fe would
show. 

The Cr 2p spectra of Fe$_{0.95}$Cr$_{0.05}$ and Fe$_{0.85}$Cr$_{0.15}$
alloys in Figs.\ \ref{fig:Fig14}b and  \ref{fig:Fig14}c show how the
concentration can affect the formation of oxide layer already at very
low pressure (10$^{-8}$ Torr). These spectra were measured using 4 keV
photon energy in which case the share of surface layer (topmost 5 nm)
has here a smaller role than in the Fe/Cr/Si spectra of Fig.\
\ref{fig:Fig14}a. In the case of higher Cr bulk concentration alloy
(Fe$_{0.85}$Cr$_{0.15}$) most of the Cr atoms are oxidized even before
any heating was carried out. The fraction of Cr$_2$O$_3$ component
of the total Cr 2p$_{3/2}$ signal area is 55 \% for
Fe$_{0.85}$Cr$_{0.15}$ and 23 \%  for Fe$_{0.95}$Cr$_{0.05}$ alloy. In
Fe$_{0.95}$Cr$_{0.05}$ the heating seems to remove part of the
Cr$_2$O$_3$ relative to unoxidized Cr making the Cr/Cr oxide signal
ratio slightly increase. Still the amount of unoxidized Cr stays
approximately the same despite the heating. Also here only the Cr
atoms were oxidized. 

\subsubsection{Cr solubility and precipitation}

While the total amount of Cr increases considerably in Fe/Cr/Si (Fig.\
\ref{fig:Fig13}) due to additional heating the share of oxidized Cr
has the highest proportion.  On the other hand, the Fe 2p spectra
measured with photon energy of 7300 eV reveal that with heating the Fe
concentration in the bulk part of the Fe-layer changes only moderately
compared to the surface region (Fig.\ \ref{fig:Fig13}), implying
similarly moderate changes for the Cr content in the bulk part of the
Fe-layer.  As the MCMD simulations show, in the Fe/Cr double layer
system chromium is dissolved in the iron layer up to the Cr solubility
limit:  6 at.\ \% at 300 K and 12 at.\ \% at 700 K (Fig.\
\ref{fig:rp_depthprofiles}).  Thus our MCMD result for the Cr
solubility limit is in line with the present HAXPES measurements of 5
and 60 minutes annealed Fe/Cr double layer showing that the Cr
concentration is lowest in the Fe film between the underlying Cr layer
and the surface region.  This demonstrates the existence of the upper
limit for the Cr solubility in Fe and therefore, in the present case,
the Fe layer acts as a retarder for the Cr diffusion from the deeper
lying Cr reservoir towards the surface and the Cr enrichment to the
surface. 

The MCMD simulations of formation of Cr rich ${\alpha}'$ precipitates
at different temperatures with different Cr concentration in
Fe$_{1-x}$Cr$_x$ systems are presented in Fig.\
\ref{fig:snapshots}. The experimental techniques used in this study
are not suitable to study precipitation in such a resolution that the
measurements could be carried out only for the areas where
precipitates occur as a comparison to the alloy areas. However, the
binding energy of Cr atoms is very sensitive to their chemical state
which is why combination of Cr bulk reference sample spectra and
information based on first principles calculations using the complete screening picture were used to fit the
Cr 2p spectra (Fig.\ \ref{fig:Fig14}). The feature labeled Cr has
higher binding energy than the component that describes the fraction of
Cr atoms in Fe-Cr alloy. This difference depends on the Cr
concentration in alloy being here approximately 0.9 eV referring to
Cr concentration around $15-30$ at.\% which cause the largest negative binding energy shift as a comparison to 100 \% Cr. The feature labelled Cr originates
from atoms that have more other Cr atoms than Fe (or O) atoms as 
nearest neighbors. In alloys where Cr concentration is as low as 5 or
15 at.\ \% the origin of this Cr bulk like structure can be either
formation of Cr rich layer under the topmost surface atom layers due
to segregation or precipitation of Cr. Cr can also have high
concentration on the grain boundaries but then it would most likely be
oxidized. According to the simulations in alloys with Cr bulk
concentration exceeding 10 at.\ \% (Fig.\ \ref{fig:snapshots}) Cr rich
${\alpha}$' precipitates can be found already at room temperature. In
the experiments the relative intensity (area of the Cr component)
increases considerably from Fe$_{0.95}$Cr$_{0.05}$ to
Fe$_{0.85}$Cr$_{0.15}$ alloy being approximately 7 and 12 \%,
respectively. This trend suggests that at least part of the Cr signal
can be caused by Cr segregated $\alpha$' phase, i.e.\  Cr precipitates in bulk, being in line with the
theory.  

However, more detailed experiments and discussion of the bilayer
sample and Fe-Cr alloys are still needed before more thorough
conclusions about precipitation and oxidation states of Cr after every
heating step can be drawn. Here the fit is used to give a better
picture how the method can be used to follow the rate of oxidation and
progression of Cr segregation and to estimate the Cr concentration
over the analyzed sample layer. When comparing the AES, XPS and HAXPES
results besides resolution and probing depth of the techniques, it is
essential to keep in mind the possible effect of sputtering on the
chemical composition of the bilayers, the different rate of oxidation
caused by differences in analyzer chamber pressures and possible
treatment and transfer of the samples prior to the measurements. All
these mentioned parameters were set to meet each other as well as
possible and the differences have been taken into account when
analyzing the results. 

\section{Conclusions}
Due to the major challenges related to the investigation of long-range properties of materials as a function of different internal and external parameters we have used in the present investigation a multi-scale and interdisciplinary approach. EMTO method has been used to investigate atomic-scale properties, MCMD method is applied for exploring the large-scale bulk phenomena, and several spectroscopic techniques have been used to study properties related to kinetics and oxidation.

The {\em ab initio} EMTO calculations predict that in the initially
homogeneous Fe$_{1-x}$Cr$_x$ the net driving force of the migration of
Cr atoms is from the second atomic layer to the bulk for low Cr alloys
($c_{\rm Cr} \lesssim 10$ at.\%) and to the surface for moderate Cr
alloys ($ c_{\rm Cr} \gtrsim$ 10 at.\%). Comparing the bulk and
surface the calculations predict the driving force to be from the
surface to the bulk ($c_{\rm Cr} \lesssim 10$ at.\%) and from the bulk
to the surface ($c_{\rm Cr} \gtrsim$ 10 at.\%), i.e. the {\em ab
  initio} simulations predict Cr containing surfaces when Cr
concentration exceeds $\sim$10 at\%. 

Monte Carlo molecular dynamics simulations predict that the formation
of Cr-rich precipitates affect the thermodynamics of the surface
layers of Fe-Cr alloy. Increasing the Cr concentration the excess
chromium goes to the precipitates. Due to the Cr surface and
segregation energies the precipitates do not reach the surface layer
and, consequently, there is no large thermodynamic driving force to
push Cr atoms to the surface. In ambient conditions oxidation of Cr
changes the picture. Surface Cr is oxidised and the oxide-alloy
interface forms a sink to the Cr of the alloy phase. One should,
however, note that these simulations only give the thermodynamic
equilibrium states. Kinetics of the alloy microstructure and surface
oxidation can be expected to have effect on the surface
structure. Simulations of Fe/Cr layer structure show the temperature
dependent upper limit of the solubility of Cr in Fe $\sim 6$ at.\% at
300 K and $\sim 12$ at.\% at 700 K. 

Experimental investigations concentrated on Cr segregation and
precipitation together with initial oxidation of Fe-Cr systems. The
segregation of Cr was nicely demonstrated in the spectra by annealing
Fe/Cr double layer and following the diffusion of Cr towards the
surface. This was done using both AES and HAXPES measurements. Besides
comparison of the intensity of Cr and Fe core level spectra in Fe/Cr
double layer and Fe-Cr alloys more detailed information of chemical
state of Cr atoms was derived from the deconvoluted Cr 2p spectra. The
fitting procedure suggested that part of the Cr atoms have bulk Cr
like structure which can be connected to the formation of Cr rich
precipitates where the nearest neighbours of Cr atoms are other Cr
atoms instead of Fe (or O) atoms. The share of this Cr signal was much
higher for Fe$_{0.85}$Cr$_{0.15}$ than for Fe$_{0.95}$Cr$_{0.05}$
alloy which is in line with the MCMD simulations. Initial oxidation of
Fe-Cr systems is provably very demanding to investigate because of
very fast Cr oxide formation in freshly cleaned samples even at very
low pressure, in the UHV of analyzer chamber of spectrometers. Cr was
the only compound that was oxidized during and after annealing and the
formation of Cr$_2$O$_3$ was verified by using the reference spectra
and fitting procedure. 

In order to further improve the understanding of the properties and phenomena related to Fe-Cr in different ambient conditions the development trends such as MC simulations using potentials which properly take account for the surface effects (e.g. enhanced magnetic moments), as well as more extensive DFT simulations using effective chemical potentials in kinetics would be advantageous. The developing HAXPES technique will also provide more selective research methods in future.

\section*{Acknowledgements}
We thank HZB for the allocation of synchrotron radiation beamtime and
Dr. Mihaela Gorgoi for all the guidance and help during the
experiments. The research leading to these results has received
funding from the European Community's Seventh Framework Programme
(FP7/2007-2013) under grant agreement N\textsuperscript{\underline{o}}
312284. Dr.\ Giovanni Bonny is acknowledged for providing the 2BEAM data for author's use. The computational facilities provided by the IT Centre for
Science (CSC) Espoo, Finland and the FGI project (Finland) are
gratefully acknowledged.

\bibliography{FeCr}

\end{document}